\documentclass[%
aip,
amsmath,amssymb,
reprint,%
]{revtex4-1}
\usepackage{color}
\usepackage{graphicx}
\usepackage{dcolumn}
\usepackage{bm}

\usepackage[utf8]{inputenc}
\usepackage[T1]{fontenc}
\usepackage{mathptmx}
\usepackage{etoolbox}
\UseRawInputEncoding
\usepackage{multirow}
\usepackage[colorlinks=true,linkcolor=blue,urlcolor=blue,anchorcolor=blue,citecolor=blue,bookmarksnumbered]{hyperref}
\makeatletter
\def\@email#1#2{%
	\endgroup
	\patchcmd{\titleblock@produce}
	{\frontmatter@RRAPformat}
	{\frontmatter@RRAPformat{\produce@RRAP{*#1\href{mailto:#2}{#2}}}\frontmatter@RRAPformat}
	{}{}
}%
\makeatother
\begin{document}
	
	\preprint{AIP/123-QED}
	
	\title{Continuous broadband Rydberg receiver using AC Stark shifts and Floquet States}
	\author{Danni Song}
\affiliation{State Key Laboratory of Quantum Optics and Quantum Optics Devices, Institute of Laser Spectroscopy, Shanxi University, Taiyuan 030006, P. R. China
	}%
 
	\author{Yuechun Jiao}
	\email{Authors to whom correspondence should be addressed: [Yuechun Jiao, ycjiao@sxu.edu.cn]}
 \affiliation{State Key Laboratory of Quantum Optics and Quantum Optics Devices, Institute of Laser Spectroscopy, Shanxi University, Taiyuan 030006, P. R. China
	}%
	\affiliation{Collaborative Innovation Center of Extreme Optics, Shanxi University, Taiyuan 030006, China}
 
	\author{Jinlian Hu}
	
	\author{Yuwen Yin}
        \author{Zhenhua Li}
 	\author{Yunhui He}
\affiliation{State Key Laboratory of Quantum Optics and Quantum Optics Devices, Institute of Laser Spectroscopy, Shanxi University, Taiyuan 030006, P. R. China
	}%
 
 	\author{Jingxu Bai}

 	\author{Jianming Zhao}
  \email{Authors to whom correspondence should be addressed: [Jianming Zhao, zhaojm@sxu.edu.cn]}
  \affiliation{State Key Laboratory of Quantum Optics and Quantum Optics Devices, Institute of Laser Spectroscopy, Shanxi University, Taiyuan 030006, P. R. China
	}%
	\affiliation{Collaborative Innovation Center of Extreme Optics, Shanxi University, Taiyuan 030006, China}
  	
	\author{Suotang Jia}
	\affiliation{State Key Laboratory of Quantum Optics and Quantum Optics Devices, Institute of Laser Spectroscopy, Shanxi University, Taiyuan 030006, P. R. China
	}%
	\affiliation{Collaborative Innovation Center of Extreme Optics, Shanxi University, Taiyuan 030006, China}

	\date{\today}
	
	\begin{abstract}
        We demonstrate the continuous broadband microwave receivers based on AC Stark shifts and Floquet States of Rydberg levels in a cesium atomic vapor cell. The resonant transition frequency of two adjacent Rydberg states 78$S_{1/2}$ and 78$P_{1/2}$ is tuned based on AC Stark effect of 70~MHz Radio frequency (RF) field that is applied outside the vapor cell. Meanwhile, the Rydberg states also exhibit Floquet even-order sidebands that are used to extend the bandwidths further. We achieve microwave electric field measurements over 1.172~GHz of continuous frequency range. The sensitivity of the Rydberg receiver with heterodyne technique in the absence of RF field is 280.2~nVcm$^{-1}$Hz$^{-1/2}$, while it is dramatically decreased with tuning the resonant transition frequency in the presence of RF field. Surprisingly, the sensitivity can be greatly improved if the microwave field couples the Floquet sideband transition. The achieving of continuous frequency and high sensitivity microwave detection will promote the application of Rydberg receiver in the radar technique and wireless communication.
	\end{abstract}
	
	
	\begin{quotation}
	\end{quotation}
	
	\maketitle
	
	In recent years, remarkable progress has been made in Rydberg atom-based electrometry~\cite{Fancher2021,yuan2023} due to their advantages in the measurement of weak microwave fields with high sensitivity, calibration-free, stability, and accuracy. An optical Rydberg electromagnetically induced transparency (EIT) and Autler-Townes (AT) splitting spectroscopy has been employed to measure the properties of electric fields, including SI-traceable standards for electric field strength~\cite{Sedlacek2012,holloway2017}, polarization measurements~\cite{Sedlacek13}, subwavelength imaging~\cite{Holloway2014} and the angle-of-arrival~\cite{RobinsonAmy2021} with a wide frequency range from DC to over 1~THz~\cite{Fan2015,jiao2017, Jau2020, Wade2017}. The sensitivity of Rydberg electrometry has been greatly improved to 55~nVcm$^{-1}$Hz$^{-1/2}$ using heterodyne technique~\cite{Jing2020} and later to 30~nVcm$^{-1}$Hz$^{-1/2}$ by adding a repumping method to enhance the EIT amplitude~\cite{Prajapati2021}. The state-of-the-art sensitivity of Rydberg microwave electrometry is improved to 5.1~nVcm$^{-1}$Hz$^{-1/2}$ by selection of higher Rydberg states~\cite{cai2023}.      
Besides the sensitivity, Rydberg atom has plentiful energy levels that cover an ultra-wide microwave frequency range. However, the EIT-AT-based Rydberg electrometry restricts the measurement of microwave fields to a series of discrete frequencies with a narrow bandwidth since it relies on resonant or near-resonant transitions between two Rydberg states. To achieve continuous-frequency microwave field detection with high sensitivity, an auxiliary microwave field resonant with an adjacent Rydberg transition is applied to achieve a tunable Rydberg resonant AT splitting~\cite{simons2021continuous,liu2022continuousfrequency,cui2023Extending}. Alternatively, the continuous frequency electric field with high sensitivity can also be achieved by utilizing AC Stark shift and in combination with heterodyne technique~\cite{meyer2021Waveguide,hu2022continuously}, where a strong far off-resonant field acts as the local field to shift the atomic energy levels~\cite{anderson2017continuousfrequency}, or by utilizing the DC Stark effect to alter the resonance frequency of the two Rydberg states, where the DC field is applied by a pair of parallel electrode plates inside the atomic vapor cell~\cite{ouyang2023continuous}. However, the electrode plates inside the cell will reflect and perturb the microwave field, which will scramble its polarization. Additionally, the measurement frequency range can be expanded using Zeeman effect to split and modify adjacent Rydberg level intervals~\cite{shi2023tunable}. However, for the mentioned approaches above, the sensitivity has a significant decrease when the resonant transition frequency is largely tuned. 

\begin{figure*}[ht]
  \centering
    \includegraphics[width=0.85\textwidth]{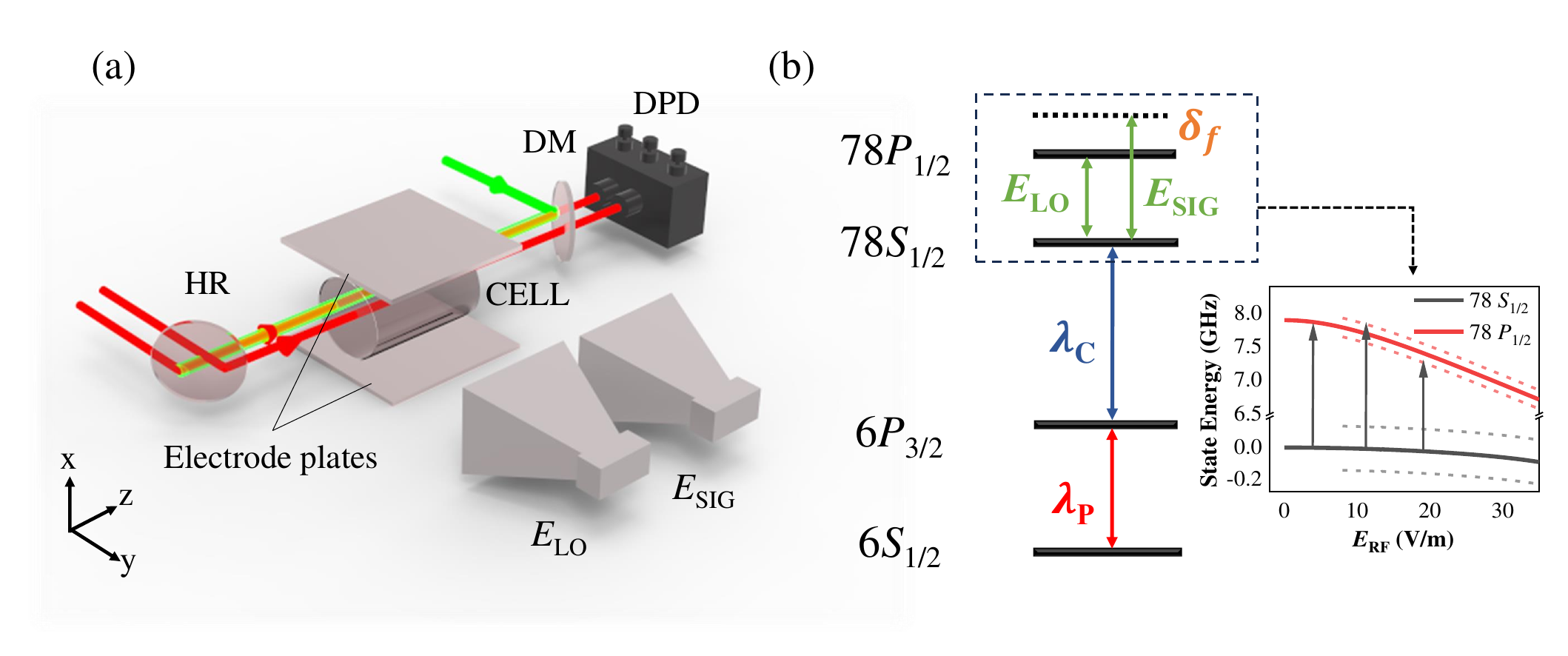}
\caption{(a) Schematic diagram of the experimental setup. An 852~nm probe laser counter-propagates and overlaps with a 510~nm Rydberg laser in the cell. The transmission of the probe and reference beam is detected by a balanced photodetector (DPD) after passing through a dichroic mirror (DM). An AC field is applied by a pair of Aluminum parallel-plate electrodes outside the vapor cell. Two microwave fields, denoted as a local oscillator (LO) field $E_{\mathrm{LO}}$ and a weak signal field $E_{\mathrm{SIG}}$, are emitted from two horn antennae. (b) Energy-level diagram for a four-level system. The probe laser is resonant with the transition of $\arrowvert6S_{1/2}, F=4 \rangle\to\arrowvert6P_{3/2}, F'=5\rangle$ and the coupling laser drives the transition of $\arrowvert6P_{3/2}, F'=5\rangle\to\arrowvert78S_{1/2}\rangle$. The LO field $E_{\mathrm{LO}}$ couples the resonant transition of 78$S_{1/2}$ and 78$P_{1/2}$, sidebands of 78$P_{1/2}$ as well, while the signal field $E_{\mathrm{SIG}}$ has a frequency difference $\delta_f$ with the LO field. The right part shows the energy shift of two Rydberg states and the generation of Floquet sidebands in the presence of an AC field.} 
\label{Fig.1}
\end{figure*}

In this work, we utilize the AC Stark effect of RF field and Rydberg Floquet states to achieve continuous broadband measurements of microwave fields with Rydberg atoms in a $^{133}$Cs vapor cell. The basic idea is that an RF field is applied to shift the resonant transition frequency of two adjacent Rydberg states and generate Floquet even-order sidebands that are coupled by the microwave field to extend the bandwidths further. Specifically, we excite the Cs ground atoms to 78$S_{1/2}$ state via a two-photon resonant EIT spectroscopy, and the microwave field couples the transition of 78$S_{1/2}$ $\to$ 78$P_{1/2}$, and sidebands of 78$P_{1/2}$ as well. The choice of both Rydberg states with $j=1/2$ is to make EIT spectrum exhibit only $m_j = 1/2$-dependent shift and sidebands in the presence of the 70~MHz RF field. We achieve detection of microwave electric field from 7.377~GHz to 6.205~GHz. The sensitivity of continuous broadband microwave field receivers is detected using the heterodyne technique. We demonstrate the sensitivity is decreased with tuning the resonant transition frequency, e.g., when the resonant frequency between 78$S_{1/2}$ and 78$P_{1/2}$ is tuned from 7.377~GHz to 6.652~GHz, the sensitivity is decreased from 280.2~nVcm$^{-1}$Hz$^{-1/2}$ to 9.566~$\mu$Vcm$^{-1}$Hz$^{-1/2}$. However, we find the sensitivity can be greatly improved if the 6.652~GHz microwave field couples the Floquet sidebands, and the sensitivity is 
1.636~$\mu$Vcm$^{-1}$Hz$^{-1/2}$, 
which is increased by a factor of 5.8. 
The use of AC field allows us to place the metal electrodes outside the cell, such the electrodes can be far away from the cell and will not perturb the microwave field. Meanwhile, the Floquet sidebands can be utilized to further extend the bandwidths and improve the sensitivity.
	
	
	The experiments are performed in a cylindrical cesium room-temperature vapor cell with 50~mm long and 25~mm diameter. The experimental setup and relevant energy levels are shown in Fig.~\ref{Fig.1}(a) and (b). Two identical 852~nm laser beams act as a probe and a reference beam that are both parallel through the cell along the z-axis. A 510~nm coupling laser counter-propagates and overlaps with the probe laser, but not the reference beam. The probe laser couples the $\arrowvert6S_{1/2},F=4\rangle\to\arrowvert6P_{3/2},F'=5\rangle$ transition with a power of 175~$\mu$W and a diameter waist of 1600~$\mu$m, while the coupling laser with a power of 56.6~mW and a diameter waist of 1800~$\mu$m drives the $\arrowvert6P_{3/2}, F’=5\rangle\to\arrowvert78S_{1/2}\rangle$ transition, such establishing EIT spectrum. The transmission of the probe and reference beams pass through a dichroic mirror (DM) and are detected by a balanced photodetector (DPD), reducing the laser intensity noise. The probe and coupling lasers keep co-linear polarization along the x-axis. A pair of Aluminum parallel-plate electrodes (size $120~{\mathrm{mm}} \times 75~{\mathrm{mm}} \times 1~{\mathrm{mm}}$) is placed outside the vapor cell with a spacing of 47~mm. A 70~MHz RF field is provided by a signal generator (Tektronix AFG3102C) using two lead wires and the electric field vector points along the x-axis. Two microwave fields, denoted as a local oscillator (LO) field $E_{\mathrm{LO}}$ and a weak signal fields $E_{\mathrm{SIG}}$, are emitted from two horn antenna (A-info LB-20180SF) and incident to the cell with co-linear polarization along the x-axis, simultaneously. The LO field frequency is resonant with the transition of $\arrowvert78S_{1/2}\rangle\to\arrowvert78P_{1/2}\rangle$, while the signal field has a 20~kHz detuning relative to the resonance transition. In the presence of the RF field, the 78$S_{1/2}$ and 78$P_{1/2}$ Rydberg states exhibit different Stark shifts due to their different polarizabilities, thereby altering the resonant transition frequency of two Rydberg states, and further the RF field induces the Floquet states of Rydberg atom~\cite{jiao2016Spectroscopy}.
	
	\begin{figure}[htbp]
  \centering    \includegraphics[width=0.45\textwidth]{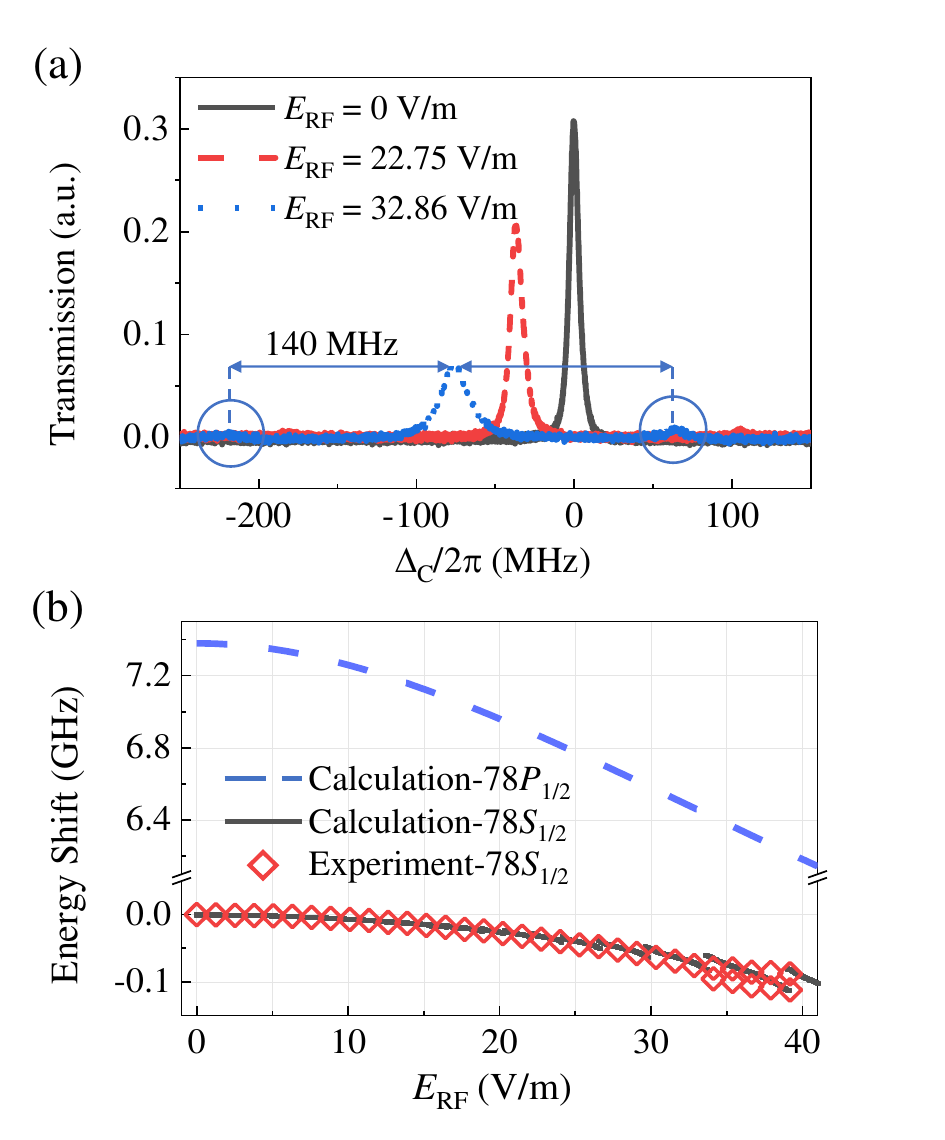}
    \caption{(a) The measured EIT spectra of 78$S_{1/2}$ state at RF field strengths of 0 (black solid line), 22.75~V/m (red dashed line) and 32.86~V/m (blue dotted line), respectively. The second-order sidebands are labeled with blue circles. (b) Stark shifts of 78$S_{1/2}$ and 78$P_{1/2}$ Rydberg states. Red diamonds represent the measured Stark shift of 78$S_{1/2}$ Rydberg state as a function of the RF field strength $E_{\mathrm{RF}}$. The black solid line and blue dashed line represent the theoretically calculated Stark shift for the 78$S_{1/2}$ and 78$P_{1/2}$ states.}   
\label{Fig.2}
\end{figure}
	
In Fig.~\ref{Fig.2}(a), we demonstrate the 78$S_{1/2}$ EIT spectra by scanning the detuning of the coupling laser $\Delta_{\mathrm{C}}$ at indicated RF field $E_{\mathrm{RF}}=0$  (black solid line), 22.75~V/m (red dashed line) and 32.86~V/m (blue dotted line). The value of $E_{\mathrm{RF}}$ represents the root-mean-square of RF field. The peak of the field-free EIT spectrum defines the 0-detuning position. It is seen that the EIT peak is red-shifted due to the AC Stark effect of RF field and there is no splitting because of only one magnetic substate $m_j=1/2$ for $S$ Rydberg states. Besides, we observe the second-order sidebands, which have band indices $N =\pm2$, separated by $\pm140$~MHz from the main peak, labeled in the blue circles for EIT spectrum at $E_{\mathrm{RF}}=32.86$~V/m. Then we perform a series of measurements such as in Fig.~\ref{Fig.2}(a) by varying the strength of RF field from 0 to 39.18~V/m in steps of 1.26~V/m and obtain the Stark shift of EIT main peak as a function of the RF field strength $E_{\mathrm{RF}}$, shown the red diamonds in Fig.~\ref{Fig.2}(b). We observe that the 78$S_{1/2}$ EIT peak exhibits two peaks due to the space of avoided crossing larger than the linewidth of the EIT at $E_{\mathrm{RF}} > 32.86$~V/m. In the following, we demonstrate our experiments at $E_{\mathrm{RF}} \leq 32.86$~V/m for simple illustration. The black solid line and blue dashed line represent the calculated DC Stark shift for the 78$S_{1/2}$ and 78$P_{1/2}$ states using the Alkali-Rydberg Calculator (ARC)~\cite{sibalic2017arc}. Since the RF field frequency is much smaller than the characteristic atomic frequency, the AC shift follows from the DC polarizability for the given 78$S_{1/2}$ state~\cite{jiao2016Spectroscopy}. The experimental RF field (x-axis) is calibrated by the theoretical calculations. We can see the Stark shift of 78$P_{1/2}$ is much larger than that of 78$S_{1/2}$, thereby altering the microwave resonant transition frequency. 

\begin{figure}[ht]
  \centering    
  \includegraphics[width=0.45\textwidth]{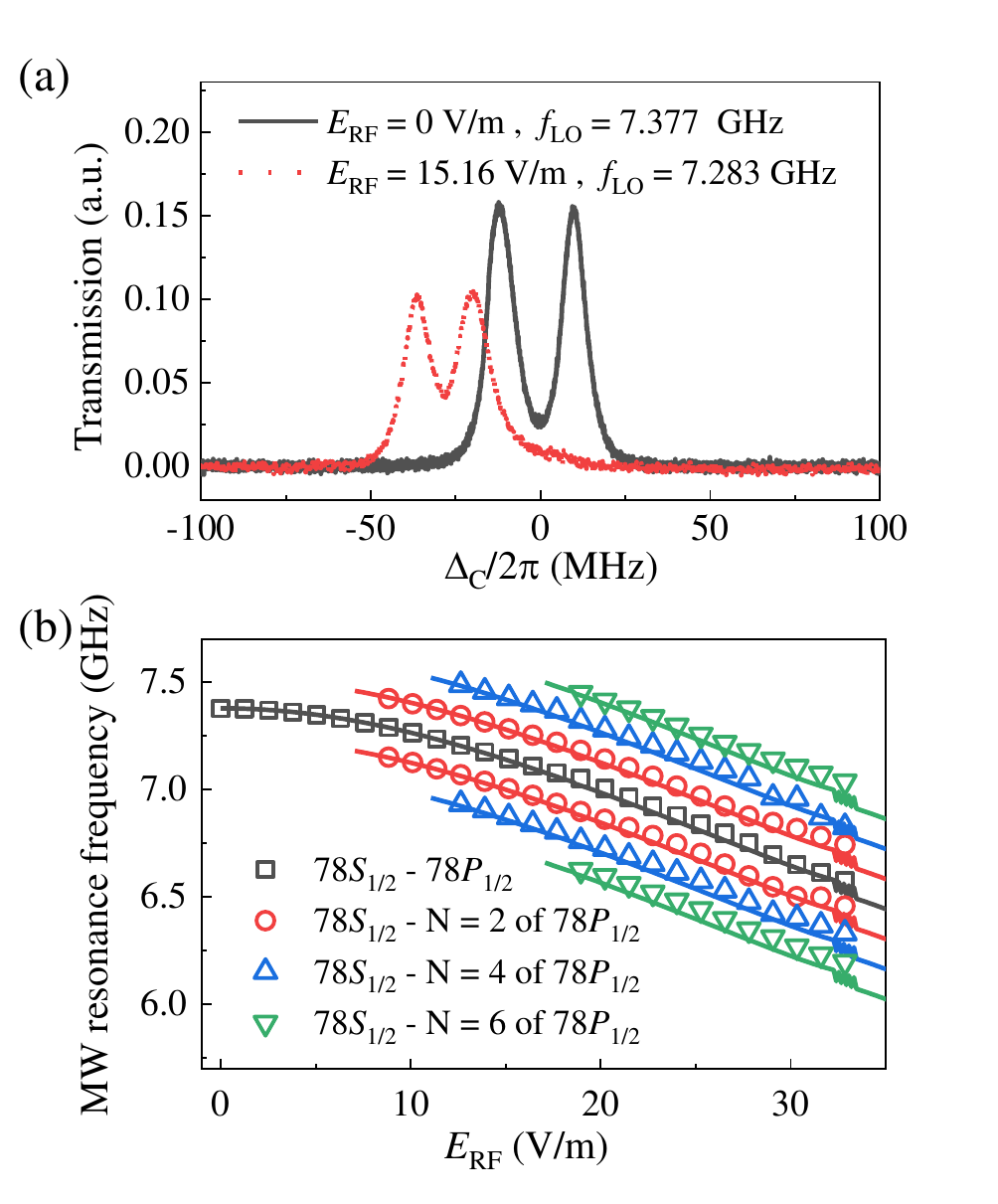}
\caption{(a) The measured EIT-AT spectra with microwave coupling the transition of 78$S_{1/2}$ $\to $ 78$P_{1/2}$ at RF electric field strengths of 0~V/m (black solid line) and 15.16~V/m (red dotted line). The corresponding microwave resonance transition frequencies are 7.377~GHz and 7.283~GHz, respectively. (b) The measured microwave resonance transition frequency as a function of the RF electric field $E_{RF}$ for the transition of 78$S_{1/2}$ $\to $ 78$P_{1/2}$  (black square) and the transition of 78$S_{1/2}$ to the Floquet sidebands of 78$P_{1/2}$ states (second-order (red circles), fourth-order (blue triangles), and sixth-order (green inverted triangles)). The black solid lines represent theoretical calculation and others are the guidelines for the even-order sidebands.}

\label{Fig.3}
\end{figure}

In Fig.~\ref{Fig.3}(a), we demonstrate two EIT-AT spectra with microwave coupling the transition of 78$S_{1/2}$ $\to $ 78$P_{1/2}$ at indicated RF field $E_{\mathrm{RF}}=0$  (black solid line), 15.16~V/m (red dotted line), where the microwave frequency is adjusted to make the splitting of two peaks symmetrically, such extracting the microwave resonant transition frequency between 78$S_{1/2}$ and 78$P_{1/2}$ are 7.377~GHz and 7.283~GHz, respectively. By measuring such EIT-AT spectra at different RF fields, we obtain the dependence of microwave resonant transition frequency on the RF field $E_{\mathrm{RF}}$, shown as the black squares in Fig.~\ref{Fig.3}(b). The results show that we continuously tune the resonant transition frequency of 
78$S_{1/2}$ $\to $ 78$P_{1/2}$ transition for the frequency range of 7.377~GHz to 6.652~GHz by varying the RF field strength. Then we use the microwave field to couple the 78$S_{1/2}$ and sidebands of 78$P_{1/2}$ and extract their transition frequency, shown as red circles (second-order sidebands), blue triangles (fourth-order sidebands) and green inverted triangles (sixth-order sidebands). By utilizing the sidebands, we further extend the measurement bandwidth to a frequency range of 7.377~GHz to 6.205~GHz. The black solid line represents the resonant transition frequency between 78$S_{1/2}$ and 78$P_{1/2}$ as a function of $E_{\mathrm{RF}}$ calculated using ARC, and other solid lines represent the guidelines for the resonant transition frequency between 78$S_{1/2}$ and even-order sidebands of 78$P_{1/2}$.


\begin{figure*}
    \centering
    \includegraphics[width=0.9\textwidth]{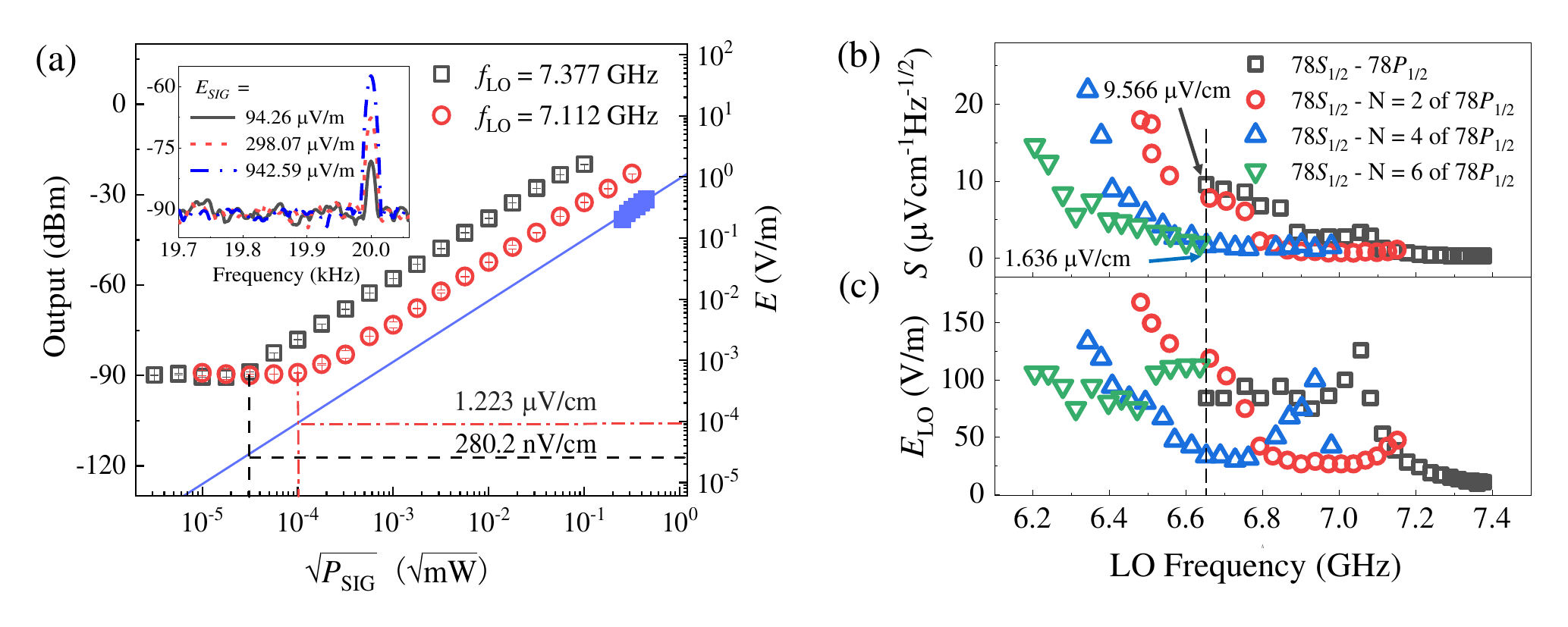}
\caption{(a) The measurement of the signal microwave field at $f_{\mathrm{LO}}=7.377$~GHz and $f_{\mathrm{LO}}=7.112$~GHz. The data were obtained with heterodyne detection (black hollow squares and red hollow circles) and EIT-AT splitting in a strong field region (blue solid squares). The blue solid line shows the calibrated electric field as a function of the microwave power $\sqrt{P_{\mathrm{SIG}}}$. The black dashed line and red dot-dashed line indicate the detectable electric fields. (b) The measurement of the sensitivity of the Rydberg receiver as a function of RF field strength $E_{\mathrm{RF}}$, corresponding microwave frequency range of 7.377~GHz to 6.205~GHz. The microwave field resonant couples the transition of 78$S_{1/2}$ $\to$ 78$P_{1/2}$ (black squares), and the transition of 78$S_{1/2}$ to Floquet sidebands of 78$P_{1/2}$ (red circles, blue triangles, and green inverted triangles). (c) The optimized LO field strength is applied for each frequency measurement in (b). }
\label{Fig.4}
\end{figure*}

Finally, we test the sensitivity of the Rydberg microwave receiver using the heterodyne technique for the 
microwave frequencies of 6.205 $\sim $7.377~GHz range, where the LO and the signal microwave fields have a 20~kHz frequency difference. When both the LO and the signal fields are incident on the Rydberg system, the transmission of the probe laser exhibits 20~kHz oscillations that are proportional to the strength of the applied signal microwave field, from which we can obtain the detectable microwave field strength by measuring the power of oscillation signals using a spectrum analyzer~\cite{Jing2020}. The inset of Fig.~\ref{Fig.4}(a) shows the measured power spectrum of the oscillation probe laser at three indicated signal fields $E_{\mathrm{SIG}}$. We can see the center frequency is located at 20~kHz and the amplitude of the signal increases with signal field strength. Here, both the probe and coupling lasers are locked to their resonant transition using an ultrastable Fabry-Perot cavity with a tunable offset-lock frequency~\cite{legaie2018SubkilohertzExcitationLasersc}. 

 In Fig.~\ref{Fig.4}(a), we demonstrate the measurement of sensitivity of the Rydberg receiver for the resonant transition between 78$S_{1/2}$ and 78$P_{1/2}$ at $E_{\mathrm{RF}}=0$ (black dashed line), 16.43~V/m (red dot-dashed line), corresponding microwave resonant transition frequency of 7.377~GHz and 7.112~GHz. During the experiments, we first measure the EIT-AT splittings to calibrate the signal field values by $E_{\mathrm{SIG}} = \frac{h}{\mu}\Delta{f}_{\mathrm{AT}}$, in which the $E_{\mathrm{SIG}}$ is proportional to the microwave power $\sqrt{P_{\mathrm{SIG}}}$. The measured electric field is shown with blue solid squares. The E-field strength at a distance of 60~cm from the cell was calculated using $E=F {\sqrt{30P\cdot g}}/{d}$, where $g$ is a gain factor of the antenna, $d$ is the distance between the horn antenna port and the center of the cell, $F$ is the cell perturbation factor. The details of the calibration process are shown in our previous work~\cite{hu2022continuously}. The calculated line is plotted in blue to calibrate the electric field strength in the weak-field region. Measurements of weak signal fields 
 as a function of microwave power $\sqrt{P_{\mathrm{SIG}}}$ using a spectrum analyzer (ROHDE$\&$SCHWARZ FSVA3013) with a resolution bandwidth of 1~Hz (measurement time of T = 1~s), shown as black hollow squares ($f_{\mathrm{LO}}=7.377$~GHz) and red hollow circles ($f_{\mathrm{LO}}=7.112$~GHz), respectively. The results show that the output of the spectrum analyzer linearly increases with the microwave power $\sqrt{P_{\mathrm{SIG}}}$. We identified the $\sqrt{P_{\mathrm{SIG}}}$ value associated with the data point exhibiting the detectable heterodyne response, then the amplitude of the microwave electric field corresponding to this specific data point is obtained by referencing the calibrated blue line, which is 280.2~nV/cm for $E_{\mathrm{RF}}=0$~($f_{\mathrm{LO}}=7.377$~GHz)  and 1.223~$\mu$Vcm$^{-1}$Hz$^{-1/2}$ for $E_{\mathrm{RF}}=16.43$~V/m ($f_{\mathrm{LO}}=7.112$~GHz), respectively.  

 Following the above process, we measure the detectable microwave electric field in a frequency range from 7.377~GHz to 6.205~GHz with microwave field coupling both the main peak and the sidebands, shown in Fig.~\ref{Fig.4}(b). 
 The black squares represent the measured sensitivity as a function of $E_{\mathrm{RF}}$ for microwave field resonant coupling 78$S_{1/2}$ $\to$ 78$P_{1/2}$ transition (zero-order sideband), the sensitivity is decreased from 280.2~nVcm$^{-1}$Hz$^{-1/2}$ to 9.566~$\mu$Vcm$^{-1}$Hz$^{-1/2}$ with tuning the transition frequency from 7.377~GHz to 6.652~GHz. The red circles, blue triangles, and green inverted triangles represent the measured sensitivity for microwave field coupling the transition of 78$S_{1/2}$ to the second, fourth, and sixth-order Floquet sidebands of 78$P_{1/2}$, respectively. We can see that the sensitivity is greatly decreased when the microwave frequency is smaller than 7.2~GHz for the 78$S_{1/2}$ $\to$ 78$P_{1/2}$ transition, while it can be substantially improved using the microwave field coupling the transition of 78$S_{1/2}$ to sidebands of 78$P_{1/2}$. For example, the case of microwave of 6.652~GHz, marked with the vertical dashed line. The sensitivity is 1.636~$\mu$Vcm$^{-1}$Hz$^{-1/2}$ for the microwave field coupling sidebands transition, which is increased by a factor of 5.8 comparison with the sensitivity of 9.566~$\mu$Vcm$^{-1}$Hz$^{-1/2}$ for the microwave field coupling main transition. The reason may be attributed to the different transition dipole moments for the microwave-induced main transition and sideband transition, which will be our further work to investigate the transition dipole moments. 
  It should be noted that before the sensitivity measurement, we first optimize the LO field strength. Fig.~\ref{Fig.4}(c) shows the optimized LO field strength applied for each frequency measurement in Fig.~\ref{Fig.4}(b), which varies from 10.58~V/m to 167.77~V/m.
	
	
 In this work, we demonstrated the continuous-frequency measurement of a weak microwave electric field over 1.0~GHz by using the AC Stark shift and Floquet states of Rydberg atoms in a room-temperature cesium vapor cell. The applied RF field (70~MHz here) is used to shift the Rydberg levels of 78$S_{1/2}$ and 78$P_{1/2}$ that exhibit different Stark shifts, altering the resonant transition frequency of two Rydberg states. Meanwhile, the Rydberg levels exhibit RF Floquet sidebands, which are used to extend the bandwidths further. The Stark shift of 78$S_{1/2}$ is measured by utilizing Rydberg-EIT spectra, and the corresponding resonant microwave transition is obtained by measuring EIT-AT splittings for both the transition of 78$S_{1/2}$ $\to$ 78$P_{1/2}$ and the transition of 78$S_{1/2}$ to sidebands of 78$P_{1/2}$. The sensitivity of Rydberg microwave receiver is demonstrated by using the heterodyne technique. The results show that the sensitivity can be greatly improved if we utilize the microwave fields to couple the RF sidebands for typical microwave frequency. Our work provides an effective method to extend the bandwidth and sensitivity by using AC Stark shift and the Floquet state of Rydberg atom. In principle, we can achieve a wider bandwidth with Rydberg levels after avoiding crossing or using a high-frequency RF field. The work here is significant for improving the Rydberg-EIT-based microwave field sensitivity and bandwidth measurement.

The work is supported by the National Natural Science Foundation of China (No. U2341211, No. 62175136, No. 12241408, and No. 12120101004); Innovation Program for Quantum Science and Technology (No. 2023ZD0300902); Fundamental Research Program of Shanxi Province (Grant No. 202303021224007); and the 1331 project of Shanxi Province.
	
	The data that support the findings of this study are available from the corresponding author upon reasonable request.
	
	\bibliography{main}
	
\end{document}